\documentclass[12pt]{book}

\usepackage[dvips]{graphicx,color}

\usepackage{makeidx,tsukuba}

\makeauthorindex

\makeindex

\begin{document}

\BookTitle{\itshape The 28th International Cosmic Ray Conference}

\CopyRight{\copyright 2003 by Universal Academy Press, Inc.}


\pagenumbering{arabic}

\chapter{A  Study Of Very Inclined Showers In The Pierre Auger
 Observatory}

\author{
M. Ave $^1$, for the Pierre Auger Collaboration \\...
{\it (1) Center for Cosmological Physics, University of Chicago,
 933 S. Ellis Av, Chicago 60637, USA}\\
}

\section*{Abstract}
The Engineering Array of the Auger Observatory has been running successfully
since 2001 and inclined showers have been recorded from the start. We have
analysed the events with zenith angle $>$ 70$^0$ recorded between May and
November 2002. The different algorithms developed to analyze these showers are
also discussed. An preliminary discussion of a reconstructed event having 20
detectors hit is presented. 
Inclined showers will be detected by the full Auger
Observatory and they will allow significant enhancement of the array
aperture. High energy events will be seen as spectacular events with 30 or 40
tanks triggered and they will provide alternative information on muon content
in air showers.

\section{Introduction}

Uncovering the origin, composition and energy spectrum of the highest energy
cosmic rays is one of the biggest challenges in astroparticle physics.  The
Pierre Auger project is the next step in the search for answers to intriguing
questions about the origin of these particles.  Before proceeding to the
construction of the full-size observatory, a subset of it, the Engineering
Array (EA), has been built and operated. A detailed description of the
 EA can be found in [1].

Ground arrays of water \v Cerenkov detectors are very sensitive to extensive air
showers at large zenith angles. In these showers the majority of the particles
reaching ground level are energetic muons and electromagnetic particles
resulting from muon interactions or decay. The rate of these showers is then
governed by their muon content and is thus sensitive to primary composition.
The understanding of these showers is also important because they constitute
the background to search for very inclined showers induced by ultra-high
energy neutrinos.
 
Inclined showers would not be very different from vertical showers except for
the fact that they develop in the upper part of the atmosphere. As a result
the electromagnetic part of the shower, produced mainly from $\pi^0$ decay, is
mostly absorbed well before the shower front reaches ground level.  However,
the muon front propagates through the atmosphere mixed with an electromagnetic
halo coming from bremsstrahlung, pair production and muon decays. This halo is
continuously generated and proportional to the number of muons, representing
less than 15\% of the signal in a \v Cerenkov tank as long as one is
sufficiently far away from the core (a few tens of meters).

 The muon energy spectrum at ground has a {\sl low energy} cutoff, caused by
 muon decay, which increases as the zenith angle rises and the average muon
 energy at ground level also increases. The {\sl high energy} part of the muon
 spectrum is also slightly different due to the rise of the pion critical 
 energy
 (the energy at which the pions are more likely to decay than to interact),
 caused by the smaller density in the upper part of the atmosphere where
 inclined showers develop.
 
These energetic muons travel making small angles to the incoming cosmic ray
direction and their trajectories are deflected by the magnetic field of the
Earth. These effects start to be important at zenith angles above 70$^0$ for
the geomagnetic latitude of the Auger Observatory. At these zenith angles the
muon density patterns at ground are very different from typical densities
measured in vertical showers that show symmetry around the shower axis. The
geomagnetic field acts as a  ``natural magnetic spectrometer''
selecting high energetic muons and deforming the symmetric pattern.

Signals of inclined showers in \v Cerenkov tanks being due to muons 
differ from those in vertical showers. Since the electromagnetic part 
is heavily suppressed inclined signals have a sharper time structure, with 
a higher rise time. The signals are proportional to the muon tracks which 
increase as the zenith angle rises. Direct light, \v Cerenkov 
 photons to fall directly onto the PMT without reflection from the tank walls,
 also increase the signal produced by inclined muons. Finally, inclined muons
 are more energetic enhancing the probability of muon interactions inside the
 tank, such as bremsstrahlung, pair production and muon nuclear interactions
 [2]. 

 In this paper we report the preliminary results of the analysis of the
inclined shower events recorded by the EA between May and November 2002.

\section{Analysis procedure and results}

The analysis of inclined showers is a 3 step procedure.  The direction of the
events are obtained by fitting the particle arrival times recorded at each
station to a plane or curved front.

To reconstruct the energy of inclined events a theoretical prediction of the
 muon density patterns is required. Geomagnetic field effects must be taken
 into account.  There are 3 independent groups working on this issue: {\it a)}
 An analytical approximation of the deflection of muons in the geomagnetic
 field based on the correlation of the muon energy with the distance to the
 shower axis, Monte Carlo simulations without magnetic field are required to
 apply this method [2]. {\it b)} Parameterizations of the number of muons in a
tank based on detailed Monte Carlo simulations with geomagnetic field 
effects. This is a two step simulation: first the hadronic shower is simulated
 in the absence of magnetic field, the position, the energy and the direction
 of the muons is kept at their production point, and in the second step the
 muons are propagated in the geomagnetic field. The second step is very fast
 in computing time, allowing to produce simulations for all possible azimuth
 angles rapidly [3]. {\it c)} An analytical approximation of the lateral and
 longitudinal distribution of muons in the absence of geomagnetic field, with
 the geomagnetic field effects implemented a posteriori in a similar way
 to method {\it a)}, [4].
 
Finally, a likelihood function is built to fit the energy of each event. Two
 different likelihood functions are being used: {\it a)} The signal in VEM is
 converted to muon numbers dividing it by the average signal produced by a
 muon at the corresponding zenith angle, Poisson statistics are then used to
 calculate the probability using the prediction of the number of muons. {\it
 b)} The other approach uses a probability density function of the signal
 produced by $k$ muons, obtained in simulations, and multiplies it by the
 Poisson probability. Details of the two methods can be found in [3], [7].
In order to account for the tank response to inclined muons with different 
energies and angles of incidence, simulations were performed using the 
GEANT [5] and SDSIM [6] packages.



We have analyzed all the events recorded between May and November 2002 with
zenith angles larger than 70$^0$ and with 5 or more stations triggered. 
In the present work we have assumed proton primaries
and the QGSJET hadronic model [8]. Different assumptions about hadronic 
models and the mass of the primary particle have a direct 
impact in the number of muons and give rise to different results. 


The left panel of Fig. 1 shows the correlation between the energy 
as reconstructed in two independent analysis. One combines 
the parameterization of the muon densities and the likelihood function 
described as {\it a)} in the previous section and 
the other corresponds to different parameterization and 
likelihood function ({\it b}). 
The agreement is encouraging.
We can define $\Delta E_0$ as the distance of each point to the line defined
 by $E_a=E_b$. Right panel of Fig. 1 shows the distribution of the ratio of
  $\Delta E_0$ to the mean energy using both algorithms. The spread of this
 distribution is 20\% and no significant offset is present.

The results presented here are preliminary. 
The agreement between the two algorithms used is encouraging but further
 checks are in progress.
 The energy of some events are very sensitive to the shape of the muon density
patterns used, specially at high energies, where the EA is small compared to
the size of the shower at ground. Fig. 2 shows an example: the density map of
a reconstructed event in the plane perpendicular to the shower axis. 
This result was obtained using method a) and displays the signals in each tank 
(converted to mean number of recorded muons). 
The position of the best-fit impact point is indicated by
a cross. The $y$-axis is aligned with the component of the magnetic field
perpendicular to the shower axis. Contour levels for 2, 5, 10 and 20 muons per
station are shown for the fit.  Events like this
will be very common in the full array and they will help to check the muon 
density patterns obtained through simulations or analytical techniques. 
 
\begin{figure}[t]
  \begin{center}

    \includegraphics[height=11.5pc]{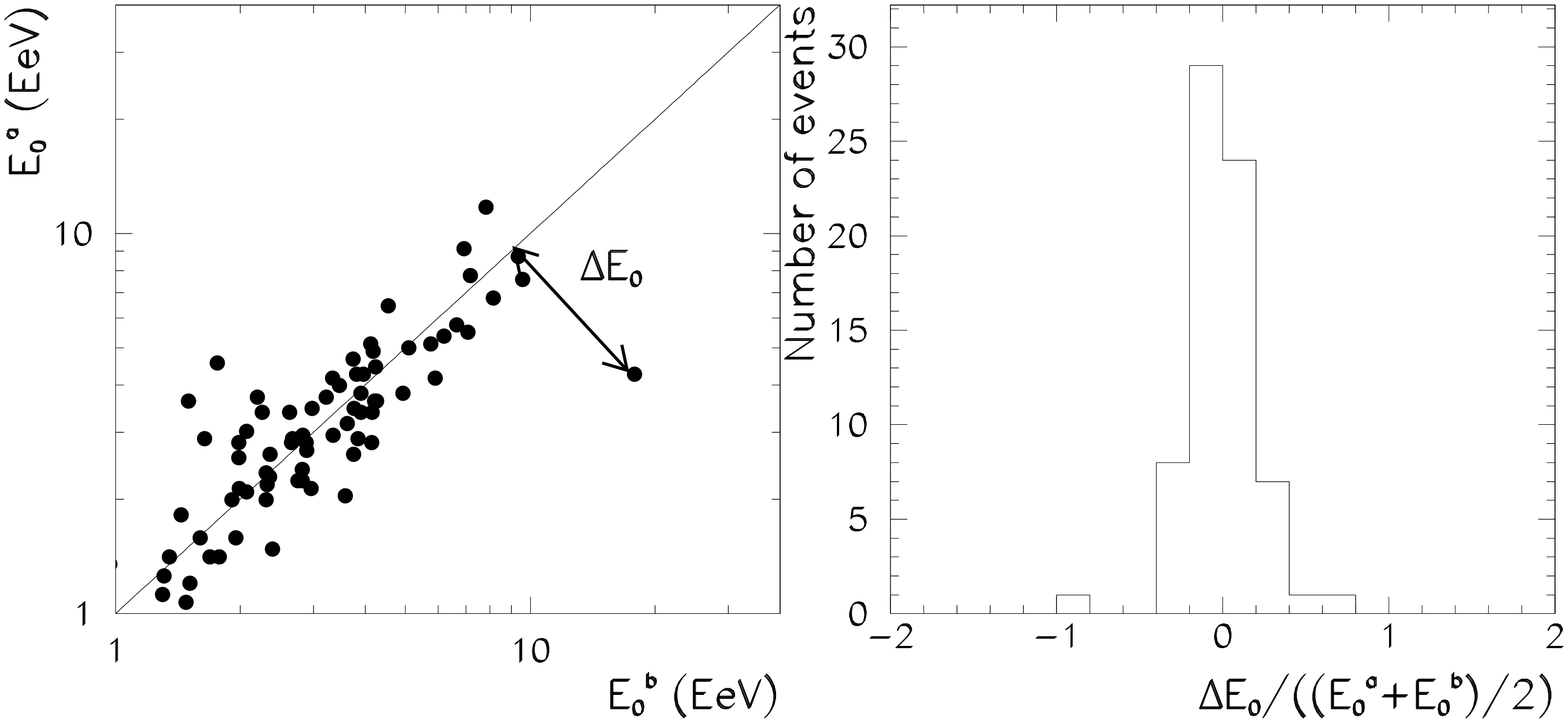}
    \includegraphics[height=12.5pc,width=14.5pc]{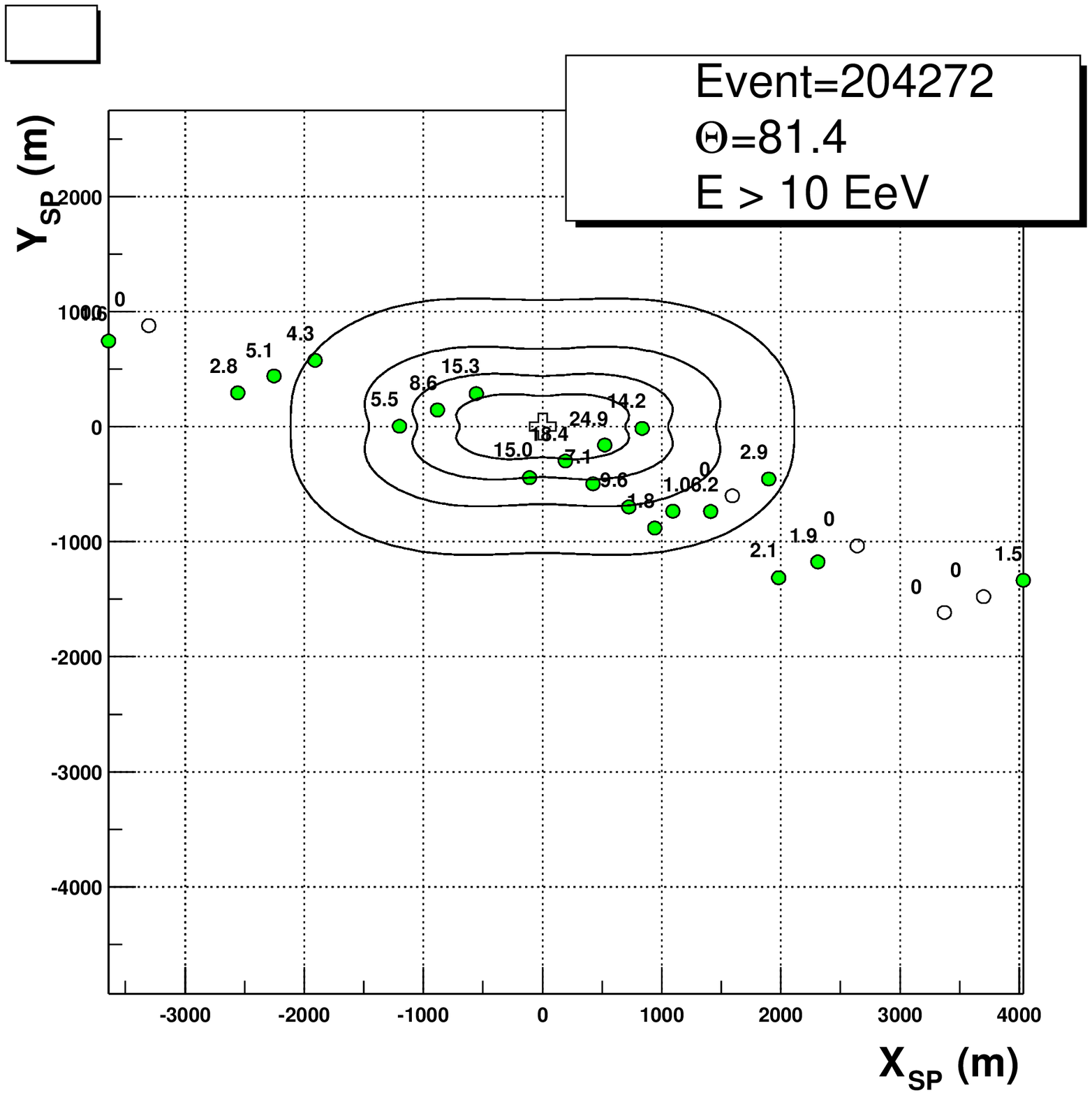}
  \end{center}
  \vspace{-0.5pc}
  \caption{Top left panel: Correlation of the
          reconstructed energy by two different algorithms, see text.
          Top right panel: Distribution of  $\Delta E_0$ to the mean 
          energy reconstructed by the two algorithms, see text.
          Bottom panel:Density map of an event, see text.}
\end{figure}

\vspace{\baselineskip}

\re

1.\ J. Abraham et al, Nucl. Inst. Met., to be published.

\re

2.\ M. Ave, R. A. V\'azquez, and E. Zas, Astropart. Phys. {\bf 14}, 91 (2000).

\re

3.\ P. Billoir, O. Deligny, A. Letessier-Savon, Internal Note GAP 2003-003.

\re

4.\ A. Dorofeev and D. Nitz, Private Communication.

\re

5.\ R. Brun {\sl et al.} GEANT, Program Library CERN (1993).

\re 

6.\ Sylvie Dagoret Campagne, Internal Note, GAP 2002-072.

\re

7.\ M. Ave, et al, Phys.Rev. D67 (2003) 043005

\re

8.\ N. Kalmykov, et al, Nucl. Phys. B {\bf 52}, 17 (1997).

\endofpaper

\end{document}